\documentclass[preprint,showpacs,amsmath,amssymb,aps,pra,floatfix,nofootinbib]{revtex4}
\usepackage{graphicx}
\usepackage{bm}
\newcommand{\be}{\begin{equation}}
\newcommand{\ee}{\end{equation}}
\newcommand{\bearr}{\begin{eqnarray}}
\newcommand{\eearr}{\end{eqnarray}}

\begin{document}
\title{ Optical Transitions at the Neutral and Charged Vacancies in Diamond }
\author{Mehdi Heidari Saani$^{*}$
, Mohammad Ali Vesaghi$^{\dag}$ 
and Keivan Esfarjani$^{\ddag}$} 
\affiliation{Department of Physics, Sharif University of Technology, Tehran,
P.O.Box: 11365-9161, Iran}
\begin{abstract}
We used the exact eigenvectors of the generalized Hubbard
Hamiltonian solution to predict the transition intensities of the
well known $GR1$ and $ND1$ transitions at the neutral and charged
vacancies in diamond. In addition to using exact eigenvectors, the
method of the calculation is more precise than already reported
calculations. The quantitative results can exlain recent
experimental data very good.
\end{abstract}
\maketitle
\newpage
\section{INTRODUCTION}
Coulson and Kearsly introduced molecule concept for vacancy
defect in diamond.$^{1}$ As the results they used the common
method for calculation the electronic structure of the molecules
in quantum chemistry i.e. Molecular orbital and Configuration
interaction.\\ In essence in this model, molecular orbital method
gives correct symmetry and spin of the electronic levels of the
vacancy and configuration interaction considers the correlation
effects of e-e interaction and gives correct sequence of the
electronic levels. Molecular orbital method has been the only way
for accounting the appropriate symmetry and spin of the vacancy
since Coulson and Kearsly model. \\Already we showed that
generalized Hubbard Hamiltonian can also take into account correct
symmetry and spin of the system in a completely different
manner.$^{2}$ This approach has origin in the solid state physics
and pay more attention to include symmetry and spin
considerations of the system in the form of a Hubbard like
Hamiltonian. This is in contrast to quantum chemistry approaches
(molecular orbital theory) which pay more attention to the
construction of the symmetry and spin adapted wavefunctions.\\
Both approaches use some parameters (8 parameters for exact
evaluation of e-e interaction) which can be obtained by theory or
semi empirical methods. These parameters can be obtained by
choosing appropriate atomic orbital function for dangling bonds
of the vacancy.\\ This evaluation is independent of the model of
Hamiltonian calculation. These parameters are the origin of the one type of
approximations which exist in the molecular model. These
parameters are the only input of the generalized Hubbard
Hamiltonian formalism and after evaluating and putting them into
the computational scheme the exact eigenvalues and eigenstates
can be obtained.\\The situation in the Coulson and Kearsly model
and its related models$^{3}$ which use molecular orbital
techniques is different. These models not only use these
approximate parameters but also apply some approximation in
constructing appropriate CI wavefunctions as the bases for
calculating the Hamiltonian.
\\Basically the coefficients of the manually constructed CI
wavefunctions in these models should be calculated by variational
principal. Also there should be some hypothesis about the most
probable configurations in the ground and excited electronic
state of the system. Additionally there should be an assumption
for truncation limit in accounting a high number of determinant
Slaters in the expansion of the CI wavefunctions.\\
These assumptions and approximations, results to a approximately
CI wavefunction for the electronic states of the vacancy in
Coulson and Kearsly and related models.$^{1,3}$  In our last
communication $^{2}$ we also explaineded some ambiguity of the
applying such CI method for this problem.\\ In summary, it is
obvious that the CI wavefunctions in molecular models are not the
exact eigenvectors of the Hamiltonian of the vacancy system.\\
Here we should note to this point that our using bases consist of
the atomic bases i.e. $a$, $b$, $c$, $d$ in contrast to Coulson
and Kearsly and all of the previous related models which use
molecular symmetric and anti symmetric bases i.e. $ a$, $t_{x}$,
$t_{y}$, $t_{z}$. These bases are more convenient in explaining
physical properties of the system. For example ionization of the
atoms of the vacancy or possibility of the exsistance of the
hunds configurations in the ground state can be verified very
easily.\\ Previously we reported $^{2}$ the probablity of finding
each electronic configuration in the exact eigenstates of the
generalized Hubbard Hamiltonian for the GR1 transition in the
neutral vacancy. (Fig.1)\\ These results and similar results for
the ground and excited states of ND1 transition are free from
approximations of the manually constructing CI wavefunctions. In
this method the origin of the error only is in the Hamiltonian parameters. \\
For verifying the validity of these new results we have used them
to calculate some experimentally observable data.
\begin{center}
\begin{figure}[t]
\includegraphics[width=5cm,height=7cm,angle=270]{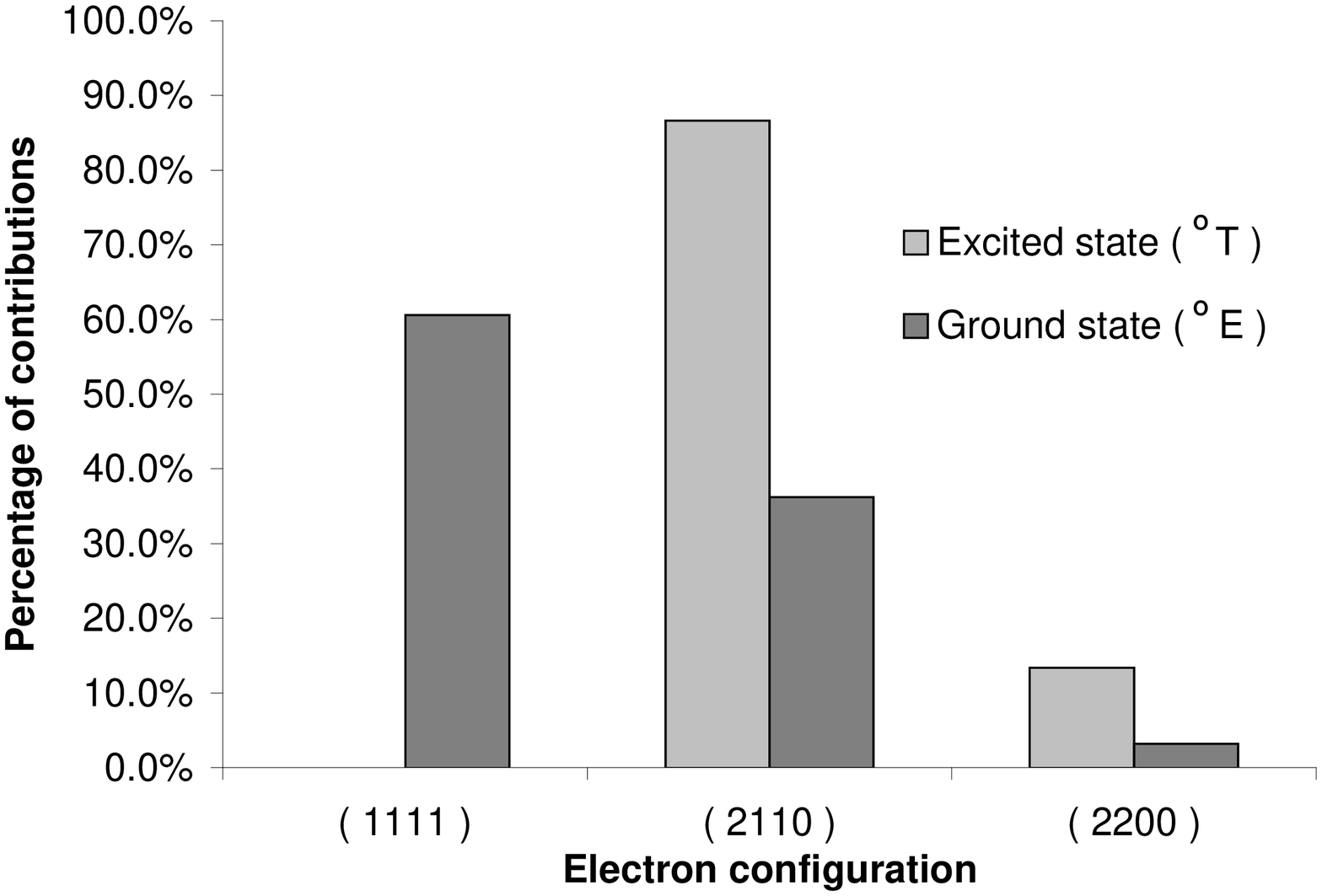}
\caption{Percentage of each possible electronic configurations in
the ground and excited states of the neutral vacancy in diamond}
\end{figure}
\end{center}
\section{MODEL }
Similar to Lowther$^{4}$ and Mainwood and Stoneham$^{5}$ we
attempted to calculate the relative intensities of the well known
GR1 and ND1 transitions in the neutral and charged vacancies in
diamond. These intensities are calculated directly with the
resultant eigenvectors of the generalized Hubbard Hamiltonina and
can be a reliable check for the validity of the physics which is
included in the exact eigenstates of the model.\\ The rate of the
electronic dipole transition between an initial state $\Psi_{i}$
and final state $\Psi_{f}$ is proportional to square of the
dipole transition amplitude i.e.
\begin{equation}
  I( i\longrightarrow f ) \propto |< \Psi_{i}\mid \overrightarrow{r}\mid
  \Psi_{f}>|^{2}
\end{equation}
Accordingly for the dipole transition from the ground states of
the $V^{0}$ and $V^{-}$ i.e. GR1 and ND1 we have two allowed
transitions which their intensities or transition rates are as
follow:
\begin{equation}
 I( GR1 ) \propto |< ^{1}E\mid \overrightarrow{r}\mid ^{1}T>|^{2}\\
\end{equation}
\begin{equation}
 I( ND1 ) \propto |< ^{4}A\mid\overrightarrow{r}\mid ^{4}T>|^{2}
\end{equation}
The eigenvectors of the generalized Hubbard Hamiltonian have
correct degeneracy due to symmetry and spin multiplicity of the
states. So we obtain two and three independent eigenvectors for
the ground and excited state of $GR1$ i.e. $^{1}E$ and $^{1}T$
respectively and we will have four and twelve independent
eigenvectors for the ground and excited states of $ND1$ i.e.
$^{4}A$ and $^{4}T$. In the end of this paper we have listed (
Table a, b) the expansion coefficients of each of these degenerate
levels on the $S_{z}$ representation of the solving Fock space
bases of the Hamiltonian. In the first quantization language this
bases are determinant Slaters which are correspond to each
electronic configuration. For simplicity, in this paper we will
call these bases as determinant.
\\In Table a, b all of
the exact eigenstates are orthonormal. An interesting point in
the Table a is high countribution of the Hunds configurations
i.e. $( 1, 1, 1, 1)$ in the ground state of $V^{0}$ and also the
absence of this configuration( model gives exactly zero! ) in the
the excited state of $GR1$ .\\ At the begining of the calculation
we note that this transition amplitudes has the same value for
the two set of the eigenvectors which transform with a unitary
matrix. So the value of the transition amplitude is the same for
$S_{z}$ or $S^{2}$ representation of the Hamiltonian eigenvectors.
\\For GR1 transition only the bases or determinant Slaters which
have $S_{z}=0$ enter into the calculation. (Table a)
 {\small{
\begin{equation}
 | ^{1}E,i>=\sum_{j=1}^{36} \alpha_{j}^{i} \varphi_{j}\hspace{1.5cm}  (i=1,2)
\end{equation}

\begin{equation}
 | ^{1}T,i>=\sum_{j=1}^{36} \beta_{j}^{i} \varphi_{j} \hspace{1.5cm}  (i=1,2,3)
\end{equation}}}
In the above equations $i$ is the degeneracy label of the states
and $\varphi_{j}$'s are the $S_{z}=0$ bases of the Fock space or
determinants. The $\alpha_{j}^{i}$ and $\beta_{j}^{i}$
coefficients are according to Table a, b. For $GR1$ transition we
have $6$ independent transition from the two degenerate $^{1}E$
states to three degenerate $^{1}T$ states. In this work we have
calculated each transition rate (intensity) separately for each
of these transitions.\\ Here we review in detail the steps of the
calculation of the intensity of one typical transition e.g.
$|^{1}E,1>$ to $|^{1}T,1>$. The amplitude of this transition from
Eq. (4), Eq. (5) is:
{\small{
\begin{equation}
 < ^{1}E,1\mid \overrightarrow{r}\mid ^{1}T,1>=\sum_{i,j=1}^{36}\alpha_{i}\beta_{j}<
\varphi_{i}\mid
\overrightarrow{r}\mid \varphi_{j}>
\end{equation}}}
In this calculation we have evaluated all of the $<\varphi_{i}\mid
\overrightarrow{r}\mid \varphi_{i}>$ terms up to $s$ order. $s$ is
the overlap integral of the two adjacent atomic orbital of the
vacancy.
\begin {equation}
s=<a|b>
\end {equation}
We have calculated the value of $s$ using appropriate Slater type
function with exponent $1.595$ as atomic orbital. The result was
$0.16 $ which agrees with the previous models.$^{1,3}$\\ These
types of amplitudes i.e. $<\varphi_{i}\mid \overrightarrow{r}\mid
\varphi_{i}>$ are direct expectation values of the dipole
$\overrightarrow{r}$ operator which arise from the same
determinant Slaters terms of the $^{1}E$ and $^{1}T$ eigenvectors
in Table a. These four electronic terms can be expanded to single
particle terms as follow.
 {\small{
\begin{equation}
 < abcd\mid \overrightarrow{r}\mid abcd>=< a\mid \overrightarrow{r}\mid a>+< b\mid
\overrightarrow{r}\mid b>+< c\mid
\overrightarrow{r}\mid c>+\\
 < d\mid \overrightarrow{r}\mid d>+O(s^{2})
\end{equation}}}
the terms which are proportional to $s^{2}$ are neglected
equation in the above equation. In addition to these direct terms
we have considered the cross terms in which the $\varphi_{i}$ and
$\varphi_{j}$ in Eq. (6) are two different determinant Slaters.
The main contribution of the cross terms comes from such
$<\varphi_{i}\mid \overrightarrow{r}\mid \varphi_{j}>$ terms in
which $\varphi_{i}$ and $\varphi_{j}$ only differ in one orbital.
These terms are proportional to $s$. For example for contribution
of the $abc\overline{d}$ and $abc\overline{c}$ determinants we
have:
 {\small{
\begin{equation}
 < abc\overline{d}\mid \overrightarrow{r}\mid abc\overline{c}>= s< a\mid
\overrightarrow{r}\mid a>+ s< b\mid \overrightarrow{r}\mid b>+ s<
c\mid \overrightarrow{r}\mid c>+\\
 < d\mid \overrightarrow{r}\mid c>+O(s^{2})
\end{equation}}}
In these expansion the sign of the terms should be considered
according to the permutation of that orbital which is different
in two determinants.\\ For simplifying the last term in Eq. (9) we
have used well known Mullikens's approximation:
 {\small{
\begin{equation}
< d\mid \overrightarrow{r}\mid c>= \frac{s}{2}(< d\mid
\overrightarrow{r}\mid d>+ < c\mid \overrightarrow{r}\mid c>)
\end{equation}}}
Now by these rules we can calculate the rate of the transition
between each degenerate ground and excited states by squaring each
transition amplitude. For simplifying the writing of the formula
we have used the abbreviation:
 \begin{equation}
 \overrightarrow{r}_{aa}= < a\mid\overrightarrow{r}\mid a>
\end{equation}
 and therefore the final amplitude of
the dipole transition will be in the form:
 {\small{
\begin{equation}
< ^{1}E,1\mid \overrightarrow{r}\mid ^{1}T,1>=
\alpha_{1}\overrightarrow{r}_{aa}+
\alpha_{2}\overrightarrow{r}_{bb}+\alpha_{3}\overrightarrow{r}_{cc}+\alpha_{4}\overrightarrow{r}_{dd}
\end{equation}}}
The $\alpha_{i}$ are obtained from coefficients which are listed
in Table a and also using value of $s= 0.16$.\\ Since the
$\overrightarrow{r}_{aa}$ , $\overrightarrow{r}_{bb}$ ,... are
not orthogonal and they have equal length, after squaring the
amplitude to find the transition rates we will have:
 {\small{
\begin{equation}
|< ^{1}E,1\mid \overrightarrow{r}\mid ^{1}T,1>|^2=(
\sum_{i=1}^{4}\alpha_{i}^{2}+
cos(\angle109.5^{o})\sum_{i,j=1}^{4}\alpha_{i}\alpha_{j})\hspace{0.2cm}
|\overrightarrow{r}_{aa}|^{2}
\end{equation}}}
\section{RESULTS AND DISCUSSIONS}
The intensities of each six and twelve allowed dipole transitions
from the two and four fold degenerate ground state of $V^{0}$ and
$V^{-}$ ( GR1 and ND1 ) are summarized in Table I, II. These
transitions are calculated by the method which was explained in
the last section. In the absence of $L-S$ coupling,
 transition between ground and excited states with different
$S_{z}$ should be zero, therefore as the results of our
calculations show, Table II should be block diagonal. The total
rates of the transitions from each of the degenerate ground state
of the $V^{0}$ and $V^{-}$ are summarized in the Table I, II.\\
\begin{widetext}
\begingroup
\squeezetable
\begin{table}[h]
\begin{center}
\begin{tabular}{c c c c c}
\hline\hline &Ground States &{$|^{1}E,1>$}&{$|^{1}E,2>$}\\
Excited states&$S_{z}$&0&0\\
\hline$|^{1}T,1>$\hspace{0.8cm}&0&0.185&0.051\\
\hline$|^{1}T,2>$\hspace{0.8cm}&0&0.025&0.092\\
\hline$|^{1}T,3>$\hspace{0.8cm}&0&0.055&0.067\\
\hline Total Rate:\hspace{0.8cm}&-&0.26&0.21\\
\hline \hline
\end{tabular}
\caption{Intensities of the transitions in the unit of
$|r_{aa}|$$(|r_{aa}|=1)$from the twofold degenerate ground state
$^{1}E$ to threefold degenerate exited state $^{1}T$ with
$S_{z}=0$ for the GR1 line.}
\end{center}
\end{table}
\endgroup
\end{widetext}
\begin{widetext}
\begingroup
\squeezetable
\begin{table}[h]
\begin{center}
\begin{tabular}{c c c c c c c }
\hline \hline &Ground states
&{$|^{4}A,1>$}&{$|^{4}A,2>$}&{$|^{4}A,3>$}&{$|^{4}A,4>$}\\ 
Excited states &$S_{z}$&-3/2&-1/2&1/2&3/2&\\
\hline$|^{4}T,1>$&-3/2&0.3&0&0&0&\\
\hline$|^{4}T,2>$&-3/2&0.31&0&0&0&\\
\hline$|^{4}T,3>$&-3/2&0.31&0&0&0&\\
\hline$|^{4}T,4>$&-1/2&0&0.14&0&0&\\
\hline$|^{4}T,5>$&-1/2&0&0.24&0&0&\\
\hline$|^{4}T,6>$&-1/2&0&0.38&0&0&\\
\hline$|^{4}T,7>$&1/2&0&0&0.23&0&\\
\hline$|^{4}T,8>$&1/2&0&0&0.075&0&\\
\hline$|^{4}T,9>$&1/2&0&0&0.44&0&\\
\hline$|^{4}T,10>$&3/2&0&0&0&0.41&\\
\hline$|^{4}T,11>$&3/2&0&0&0&0.32&\\
\hline$|^{4}T,12>$&3/2&0&0&0&0.0094&\\
\hline Total Rate:&-&0.92&0.76&0.74&0.74&\\
\hline \hline
\end{tabular}
\caption{Intensities of the transitions in the unit of
$|r_{aa}|(|r_{aa}|=1$)from the fourfold degenerate ground state
$^{4}A$ to twelvefold degenerate exited state $^{4}T$ with
available $S_{z}$ for the ND1 line.}
\end{center}
\end{table}
\endgroup
\end{widetext}
As it is seen from the tables the total rates of the transition
from each degenerate ground state are nearly equal. Physically
this sounds good. We expect that in an spectroscopic experiment
which does not distinguish the degeneracy of the ground and
excited states, one can not create a change in the relative
population of the balanced degenerate ground states by
illumination of light.$^{6}$\\ For obtaining the effective
transition rate between degenerate ground states we should
average over the the total transition rates.$^{6}$ This arises
from this point that the probability of the existence of the
system in each of the degenerate states is equal. In other hand
the number of the electrons which populate the degenerate ground
states are equal.\\ Total transition rates of the GR1 and ND1
degenerate ground states are listed in the last line of the Table
I, II.\\ From Table I we obtain:
 {\small{
\begin{equation}
  |< ^{1}E\mid \overrightarrow{r}\mid ^{1}T>|^{2}=\frac{1}{2} \sum_{i,j=1}^{2,3}|<
^{1}E,i\mid \overrightarrow{r}\mid
  ^{1}T,j>|^2=0.23|\overrightarrow{r}_{aa}|^{2}
\end{equation}}}
and from Table II we obtain:
 {\small{
\begin{equation}
  |< ^{4}A\mid \overrightarrow{r}\mid ^{4}T>|^{2}=\frac{1}{4} \sum_{i,j=1}^{4,12}|<
^{4}A,i\mid \overrightarrow{r}\mid
  ^{4}T,j>|^2=0.79|\overrightarrow{r}_{aa}|^{2}
\end{equation}}}

With these values we obtain the relative intensities of the GR1
and ND1 absorption lines:
\begin{equation}
\frac{I( ND1 )}{I( GR1 ) }=3.4
\end{equation}
This is in excellent agreement with the recent experimental work
on the GR1 and ND1 transitions intensities.$^{7, 8, 9}$ We have
used these values from Davies $^{10}$ who has calibrated them. The
experimental data shows that the relative intensities of the ND1
and GR1 is equal to 3.25$^{7, 8}$or 4$^{7, 9}$. Our theoretical
value is also consistant with the experimental error which is
reported more recently.$^{11}$  The error of our calculation
which arises from averaging over the total transition rates are
10 percent for ND1 and 11 percent for GR1. These errors should be
due to approximations in obtaining the Hamiltonian parameters.
This means that if we were able to estimate exact atomic orbitals
of the vacancy and also Hamiltonian parameters
then we expect that the total rates of the transition be equal. (last line of Tables I, II)\\
One important question in this problem has been the importance of
the overlap effect of the vacancy orbitals. We have recalculated
the intensities with the neglecting the overlap of the orbitals
i.e. $s=0$. The results for the GR1 intensity is $0.23$ which is
same as Eq. (14) and for ND1 the result is $0.87$, higher than the
value of the Eq. (15). By this assumption we reach to $3.8$ for
the relative transitions. The independent nature of the GR1
transition from the overlap parameter $s$ comes from the high
contribution of the Hunds state in the ground state. As it was
shown in Eq. (8) the transition amplitude of these type
determinants is independent from $s$. However the value of the
$s$ depends on the choosing of the atomic orbitals of the vacancy
so by changing it the Hamiltonian parameters also should be
changed. If we were able to include semiempirical nature of the
generalized Hubbard Hamiltonian parameters to the shape of the
atomic orbitals then we can be sure that the resultant overlap
value is more valid. However since the Hamiltonian parameters are
from Coulson and Kearsly calculation, the value of the $s$ in our calculation are choose to be $0.16$.\\
Our model is more similar to Mainwood and Stoneham model since
their model does not need to detail form of atomic orbital i.e.
calculation of the $|\overrightarrow{r}_{aa}|$. Also they have
attributed the GR1 and ND1 transition only to extra atomic
transitions. They have neglected the overlap of the vacancy
orbitals ($s=0$) maybe due to the order of the error which exist
in the approximated wavefunctions of the Lannoo model.$^{3}$
\section{CONCLUSION}
We applied the exact wavefunctions of the generalized Hubbard
Hamiltonian to the transition intensities of the well known
optical absorption lines of the diamond vacancies. This
calculation can successfully explain recent experimental data.
These results clearly show the advantage of using exact
wavefunctions of the generalized Hubbard Hamiltonian with respect
to previous CI wave functions. For comparing the different
wavefunctions of the two model it is useful to compare the number
of determinants which comes into the calculation of the
transition rates. For example CI only predicts $1\times1$
determinants for ND1 ground and excited states however here we
considered $8\times8$ determinants for the $S_{z}=\pm\frac{3}{2}$
and $48\times48$ for $S_{z}=\pm\frac{1}{2}$ as are reported in
Table b. Previous molecular orbital approaches were not able to
consider different forms of the degenerate eigenvectors of the
Hamiltonian and this can be another reason fort their poor
quantitative results.\\ The results of the calculation show that
the electronic configurations of the ground and excited states of
GR1 ( Fig. 1) can explain quantitatively experimental data. This
means that the ground state of the neutral vacancy mainly comes
from the Hunds rule( 60 percent). High contribution of the Hunds
states in the
ground state of the neutral vacancy physically was expected.\\
New results of present model show that generalized Hubbard
Hamiltonian has potential to go toward a quantitative
understanding of the point defects problem in diamond.
\section{Acknowledgment}
We would like to thanks Prof. J. E. Lowther, Prof. A. M. Stoneham,
\\Dr. J. Goss and Prof. J. M. Baker for their comments and
discussions.
\\\\ $^{*}$e-mail: Heydaris@mehr.sharif.edu\\
   $^{\dag}$e-mail: Vesaghi@sharif.edu
\\ $^{\ddag}$e-mail: K1@sharif.edu

\begin{center}
\begin{figure}[h]
\includegraphics[width=15cm,height=25cm,angle=0]{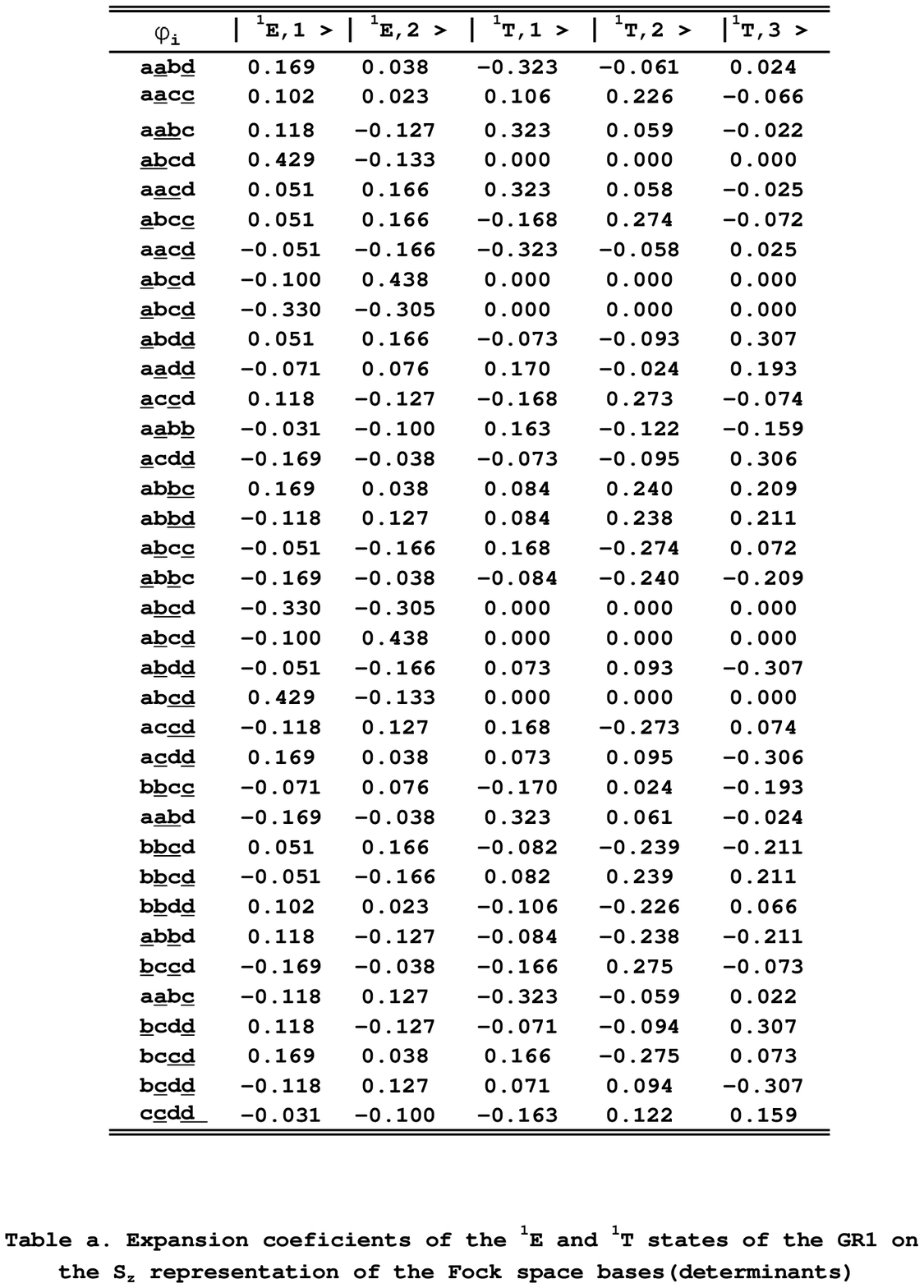}
\end{figure}
\end{center}
\begin{center}
\begin{figure}[h]
\includegraphics[width=15cm,height=25cm,angle=0]{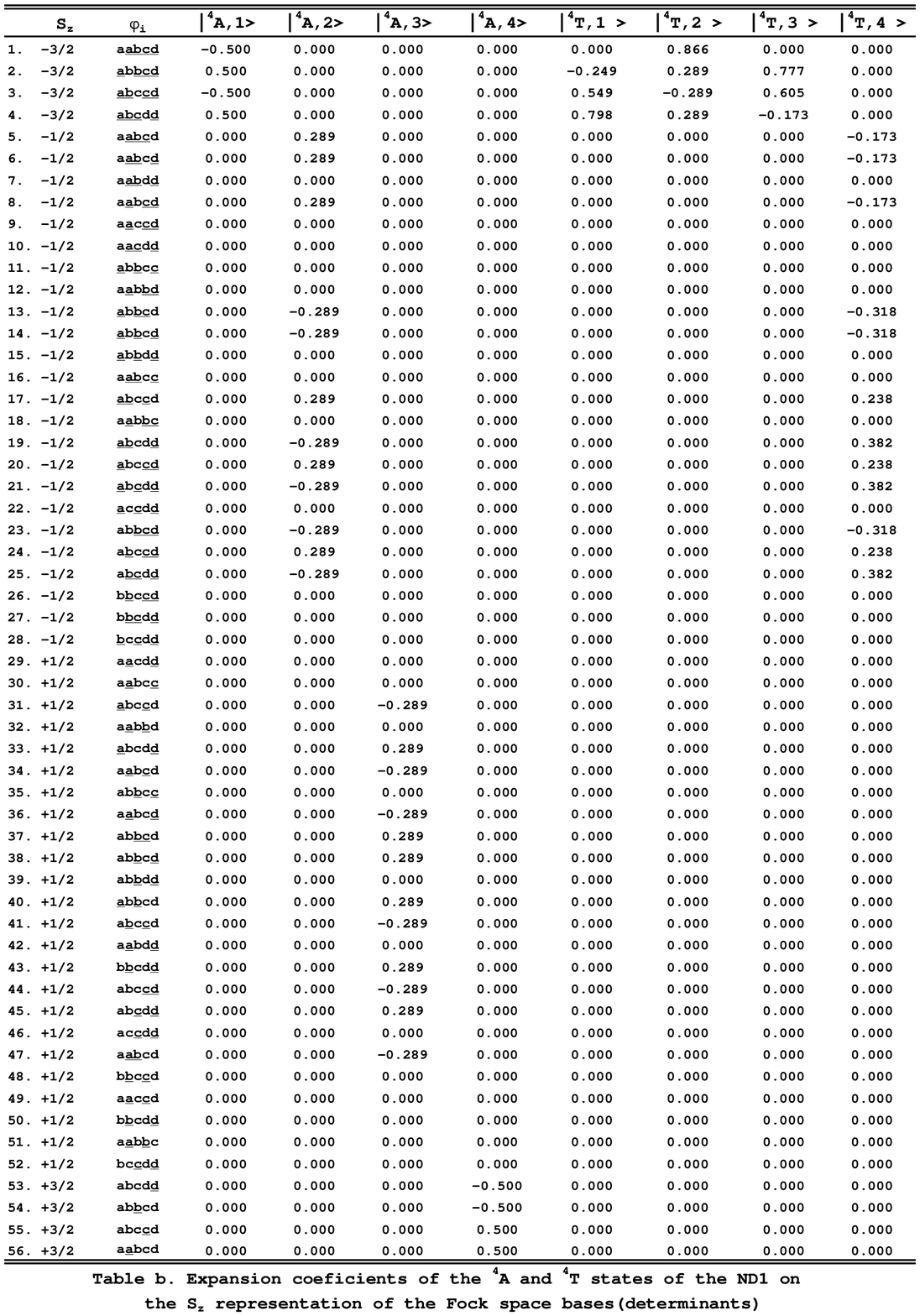}
\end{figure}
\end{center}
\begin{center}
\begin{figure}[h]
\includegraphics[width=15cm,height=25cm,angle=0]{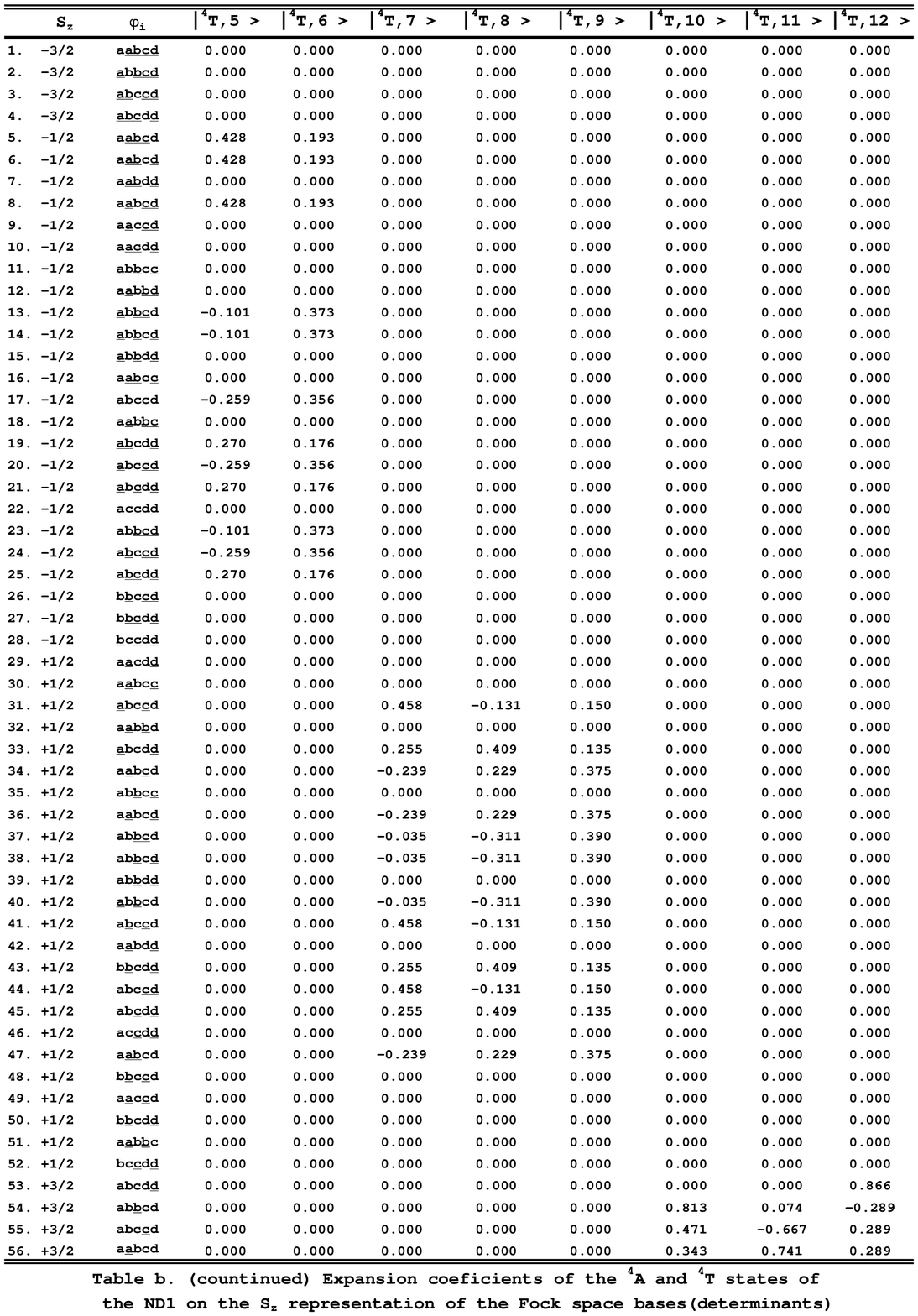}
\end{figure}
\end{center}
\end{document}